# Geometric nested sampling: sampling from distributions defined on non-trivial geometries

**Kamran Javid**[1, 2]

**1** Kavli Institute for Cosmology Cambridge, Madingley Road, Cambridge, CB3 0HA, United Kingdom **2** Astrophysics Group, Cavendish Laboratory, JJ Thomson Avenue, Cambridge CB3 0HE, United Kingdom

## Abstract

Metropolis Hastings nested sampling evolves a Markov chain, accepting new points along the chain according to a version of the Metropolis Hastings acceptance ratio, which has been modified to satisfy the nested sampling likelihood constraint. The geometric nested sampling algorithm I present here is based on the Metropolis Hastings method, but treats parameters as though they represent points on certain geometric objects, namely circles, tori and spheres. For parameters which represent points on a circle or torus, the trial distribution is "wrapped" around the domain of the posterior distribution such that samples cannot be rejected automatically when evaluating the Metropolis ratio due to being outside the sampling domain. Furthermore, this enhances the mobility of the sampler. For parameters which represent coordinates on the surface of a sphere, the algorithm transforms the parameters into a Cartesian coordinate system before sampling which again makes sure no samples are automatically rejected, and provides a physically intuitive way of the sampling the parameter space.

## Bayesian inference and nested sampling

For a model $\mathcal{M}$ and data $\mathcal{D}$, we can obtain model parameters (also known as input or sampling parameters) $\theta$ conditioned on $\mathcal{M}$ and $\mathcal{D}$ using Bayes' theorem:

$$\Pr(\theta|\mathcal{D},\mathcal{M}) = \frac{\Pr(\mathcal{D}|\theta,\mathcal{M})\Pr(\theta|\mathcal{M})}{\Pr(\mathcal{D}|\mathcal{M})},$$

where $\Pr(\theta|\mathcal{D},\mathcal{M}) \equiv \mathcal{P}(\theta)$ is the posterior distribution of the model parameter set, $\Pr(\mathcal{D}|\theta,\mathcal{M}) \equiv \mathcal{L}(\theta)$ is the likelihood function for the data, $\Pr(\theta|\mathcal{M}) \equiv \pi(\theta)$ is the prior probability distribution for the model parameter set, and $\Pr(\mathcal{D}|\mathcal{M}) \equiv \mathcal{Z}$ is the Bayesian evidence of the data given a model $\mathcal{M}$. The evidence can be interpreted as the factor required to normalise the posterior over the model parameter space:

$$\mathcal{Z} = \int \mathcal{L}(\theta) \pi(\theta) \, \mathrm{d}\theta.$$

.

Nested sampling (Skilling & others, 2006) exploits the relation between the likelihood and prior volume to transform the integral which gives the evidence as a one-dimensional integral. The prior volume $X$ is defined by $\mathrm{d}X = \pi(\theta)\,\mathrm{d}\theta$, thus $X$ is defined on $[0,1]$ and we can say:

$$\mathcal{Z} = \int_0^1 \mathcal{L}(X) \mathrm{d}X.$$

The integral extends over the region(s) of the parameter space contained within the iso-likelihood contour $\mathcal{L}(\theta) = \mathcal{L}$ (see Javid (2019a) for a proof of the equivalence of both evidence integrals).

Nested sampling has been used extensively within the astrophysics community for both computing the posterior distributions of parameters of interest and for model comparison (see e.g. (Farhan Feroz et al., 2009; Javid, 2019a; Javid, Olamaie, et al., 2018; Javid, Perrott, et al., 2018; Javid et al., 2019; Perrott et al., 2019) which all use data from the Arcminute MicroKelvin Imager (Hickish et al., 2018) to solve Bayesian inference problems). Two of the most popular nested sampling algorithms are MultiNest (F Feroz et al., 2009) and PolyChord (Handley, Hobson, & Lasenby, 2015; Handley et al., 2015). The former of these approximates the iso-likelihood contours as hyperellipsoids in the unit hypercube in which it samples. PolyChord performs coordinate transformations such that the resultant space is easier to navigate.

A comparison between the performance of the geometric nested sampler and MultiNest is presented in Javid (2019b). Figures 6 - 9 of that paper show their performance on a toroidal model, while Figure 11 shows their relative performance on a complex spherical toy model.

## Summary of the geometric nested sampler

Here I present the geometric nested sampling algorithm, as described in Javid (2019b). The geometric nested sampling algorithm is based on the Metropolis Hastings nested sampling algorithm presented in Feroz & Hobson (2008), which at each iteration, evolves an Markov Chain Monte Carlo (MCMC) chain to find the next livepoint (point within the iso-likelihood contour). Each point in the chain is sampled from a trial distribution, and whether that point is accepted or not, depends on the value of the Metropolis Hastings ratio (Chib & Greenberg, 1995). In the case that the point is not accepted, the next point in the chain is taken to be the point from the previous step.

One key issue with Metropolis Hastings nested sampling is that at each nested sampling iteration, if too many trial points are rejected, then the livepoints will be highly correlated with each other after a number of nested sampling iterations. To prevent this, one must sample a large number of trial points in order to increase the number of acceptances and decrease the auto-correlation of the trial point chain.

This method can be problematic if computing the likelihood is computationally expensive. One particular case in which the sampled point is guaranteed to be rejected, is if the point lies outside of the domain of $\mathcal{P}$ (support of $\pi$). Of course, this can be avoided by adapting $q(\theta_\mathrm{t}|\theta_l)$ so that it is truncated to fit the support of $\pi$, but in high dimensions this can be tedious, and inefficient in itself. Hence one desires an algorithm which does not sample outside the support of $\pi$, without having to truncate $q$.

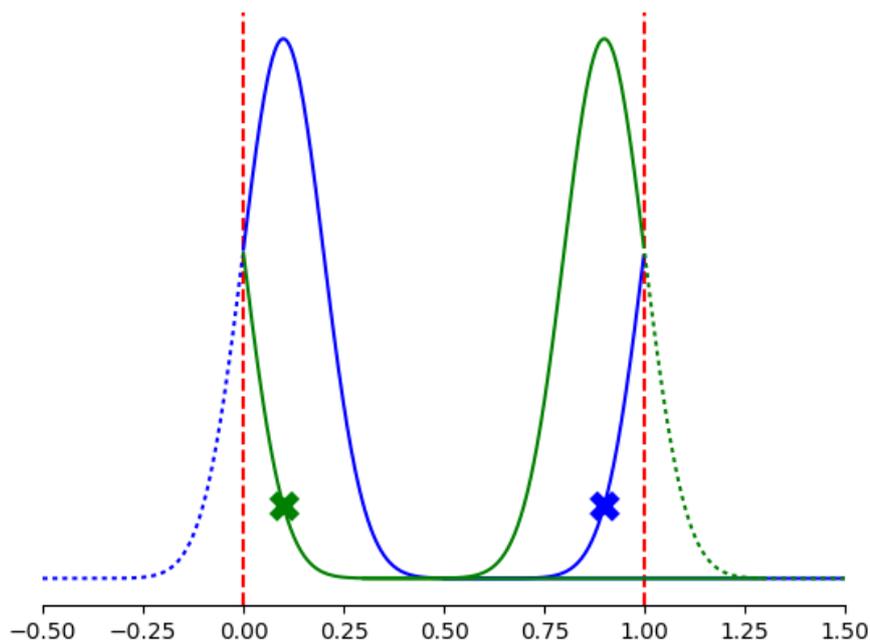

**Figure 1:** Wrapped trial distribution for circular parameters. Dotted red lines denote edge of posterior domain. Green and blue curves are candidate trial distributions. The dotted part of these curves are the parts of the distribution which would sample outside of the posterior domain if a wrapped trial distribution was not used. The crosses are examples of how a trial point could be sampled when the wrapped trial distributions are used.

Another issue which most sampling algorithms are subject to occurs when the modes of the posterior distribution are far away from each other in $\theta$ space, e.g. when they are at "opposite ends" of the domain of $\pi$. In the context of nested sampling this can result in one or more of the modes not being sampled accurately, particularly in the case of low livepoint runs. Thus a sampling algorithm should be able to efficiently manoeuvre between well separated modes which lie at the "edges" of $\pi$'s support.

Specific to the geometric nested sampler, is the way it treats parameters that exhibit certain geometrical properties. For example, for periodic parameters (e.g. time or angle) the trial distribution used in the MCMC evaluation is "wrapped" around its domain as shown in Figure 1. A wrapped trial distribution such as this one ensures that samples are never automatically rejected for being outside of the domain of the posterior. Furthermore, any modes occurring near the edge of the domain are still "close to the other end" of the domain, and so can be sampled more efficiently. For parameters which exhibit spherical properties (e.g. polar and azimuthal angles $\theta$ and $\phi$), the geometric nested sampler uses another trick. Instead of considering their joint rectangular sampling space with area $\pi \times 2\pi$, the algorithm transforms their values into their three-dimensional Euclidean representation $(x, y, z)$, and samples from a spherical Gaussian trial distribution. Any sampled points which do not lay on the original sphere described by $\theta$ and $\phi$ are projected back onto the sphere before being converted back to their angular forms, which ensures no trial points outside of the domain are sampled. This idea is illustrated in Figure 2: at iteration $l$ of the algorithm we consider a point $(\theta_l, \phi_l)$ as a point in the MCMC chain. This point can be transformed into its Cartesian representation $(x_l, y_l, z_l)$, which forms the centre of the trial distribution from which a candidate point is sampled. In general the point sampled from the trial distribution, $(x', y', z')$, will not lay on the surface of the sphere. Thus the point is projected back onto the sphere to give $(x_t, y_t, z_t)$, which can be converted back into angular coordinates to give the trial point $(\theta_t, \phi_t)$.

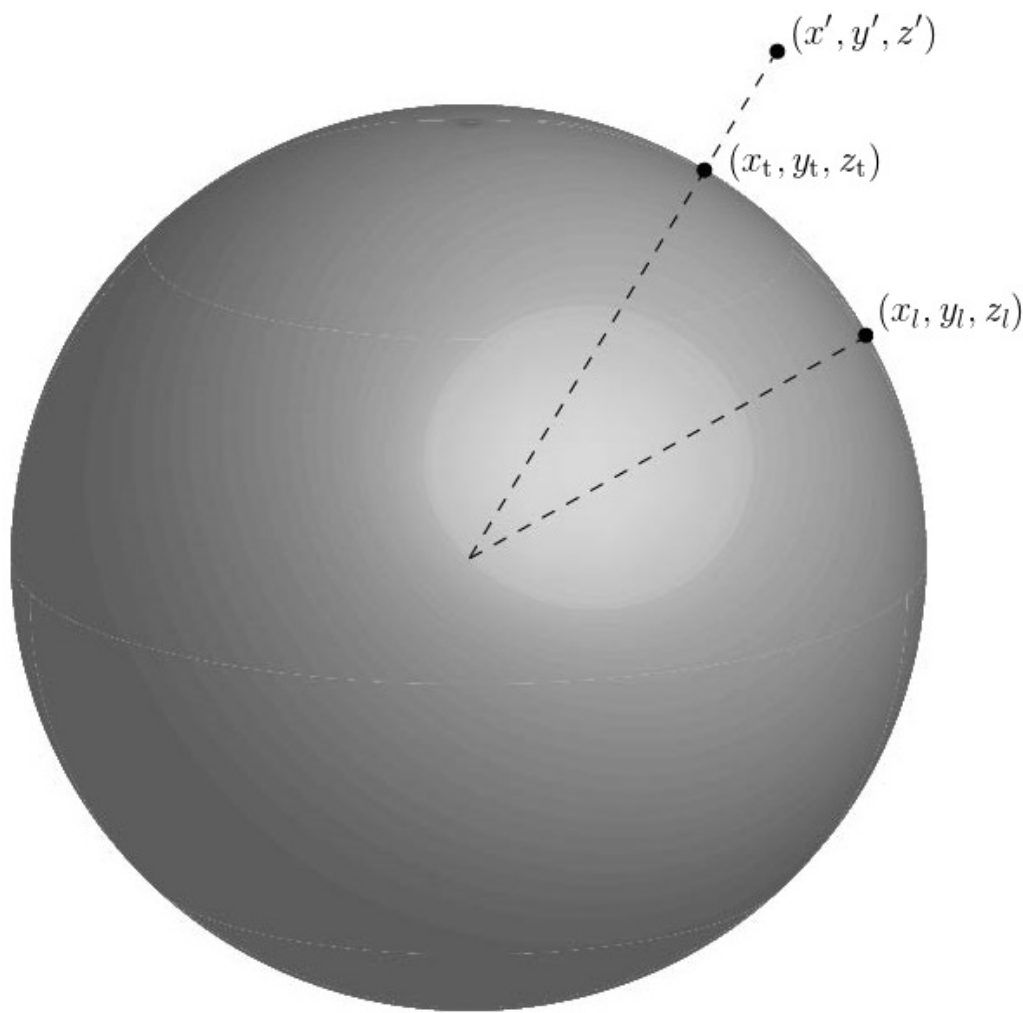

**Figure 2:** Sampling spherical parameters by using their Euclidean interpretation. $(x_l, y_l, z_l)$ represents the centre of the trial distribution, from which point $(x', y', z')$ is sampled. Since this point does not lay on the sphere, it is projected back onto it (to give point $(x_t, y_t, z_t)$) before it is converted back to its angular representation.

# Acknowledgements

The author wishes to thank Will Handley for his contribution to the project associated with this algorithm.